%% file: Template.tex
\documentclass{sig-alternate-05-2015}

\usepackage{epigraph}

\usepackage{hyperref}
\hypersetup{
    colorlinks=true,
    citecolor=blue,
    linkcolor=blue,
    filecolor=blue,      
    urlcolor=blue,
} 

\include{commands}

\usepackage{csquotes}
\usepackage{balance}  % to better equalize the last page

\begin{document}

\remove{
\CopyrightYear{2017} 
\setcopyright{acmcopyright}
\conferenceinfo{WSDM 2017,}{February 06-10, 2017, Cambridge, United Kingdom}
\isbn{978-1-4503-4675-7/17/02}
\acmPrice{\$15.00}
\doi{http://dx.doi.org/10.1145/3018661.3018709}
}

\clubpenalty=10000 
\widowpenalty = 10000

\title{Fun Facts: Automatic Trivia Fact Extraction from Wikipedia}
%\dnote{are we ok with the name? David?} \dmtsurel{Is ``Extraction'' better than ``Retrieval''?} 
%\ig{We're not really doing retrieval. Other options are ``Inference'' or ``Detection''}}

%\\ (working title \ig{maybe: Fun Facts: Extracting Trivia Facts from Wikipedia)} 
%\dnote{no need to stress categories in the title, I think} \\ \ig{Probably not in the title, but in the abstract}
% title

\remove{
\author{ 
   \large Anonymized Authors \\
   \normalsize  Affiliation  \\[-3pt]
    \normalsize Address \\[-3pt]
    \normalsize	email Address}
}

\numberofauthors{4} 
\author{
\alignauthor
David Tsurel\\
       \affaddr{The Hebrew University of Jerusalem}\\
       \email{dmtsurel@cs.huji.ac.il}
% 2nd. author
\alignauthor
Dan Pelleg\\
       \affaddr{Yahoo Research, Israel}\\
       \email{pellegd@acm.org}
% 3rd. author
\alignauthor Ido Guy\\
       \affaddr{Yahoo Research, Israel}\\
       \affaddr{Ben-Gurion University of the Negev}\\
       \email{idoguy@acm.org}
\and  % use '\and' if you need 'another row' of author names
% 4th. author
\alignauthor Dafna Shahaf\\
       \affaddr{The Hebrew University of Jerusalem}\\
     \email{dshahaf@cs.huji.ac.il}
}

\maketitle

\input{0-abstract}
%\input{keywords}
%\keywords{trivia; fun facts; factoids; serendipity; surprise}
\input{1-intro}
\input{2-problem.tex}

\input{3-algorithm}

\input{4-evaluation}
\input{5-discussion}

\input{6-rw}
\input{7-conclusions}

\newpage
\balance
%{\small
%\balance{}

\bibliographystyle{abbrv}
\bibliography{connect}
%\vspace*{1mm}
%}

\end{document}

%% file: commands.tex
\newcommand{\remove}[1]{}

\usepackage{bbding}

\usepackage{amssymb}
\usepackage{amsmath}
\usepackage{latexsym}
\usepackage{graphicx}

\usepackage{xspace}
\usepackage{dsfont}
\usepackage{graphicx}  
\usepackage{verbatim}
\usepackage{appendix}
\usepackage{framed}
\usepackage{color}

\usepackage{tikz}
\definecolor{shade}{HTML}{D4D7FE}
\definecolor{shadeblue}{HTML}{7ABDFF}
\definecolor{shadegreen}{HTML}{7AFF7A}
\definecolor{shadegreen2}{HTML}{33FF66}
\definecolor{shadewheat}{HTML}{F5DEB3}
\definecolor{shadeyellow}{HTML}{FFFF7A}
\definecolor{shadeorange}{HTML}{FFBD7A}
\definecolor{shadepurple}{HTML}{BD7AFF}
\usepackage{booktabs}

\newcommand{\xhdr}[1]{\vspace{1mm}\noindent{{\bf #1.}}}
\newcommand{\article}{\ensuremath{a}}
\newcommand{\catg}{\ensuremath{C}}
\newcommand{\wordsim}{\ensuremath{\sigma}}
\newcommand{\artsim}{\ensuremath{\sigma}}
\newcommand{\artcatsim}{\ensuremath{\sigma}}
\newcommand{\surprise}{\ensuremath{\textit{surp}}}
\newcommand{\cohesive}{\ensuremath{\textit{cohesive}}}
\newcommand{\interest}{\ensuremath{\textit{trivia}}}
\usepackage[framemethod=TikZ]{mdframed}

\definecolor{shadecolor}{rgb}{.6,.6,.6}
 {\vspace*{0.5\baselineskip}
\begin{center}\begin{minipage}{0.85\textwidth}\begin{shaded}}%
\usepackage[tight]{subfigure}

\definecolor{Emerald}{rgb}{0.1,0.8,0}

\numberwithin{equation}{section}

\newcommand{\blockcomment}[1]{ }

\usepackage{algorithm}
\usepackage[noend]{algpseudocode}

\makeatletter
\def\BState{\State\hskip-\ALG@thistlm}
\makeatother

\newbox\subfigbox


\usepackage{fancybox}

\usepackage{wrapfig}
\usepackage{bbm}
\usepackage{graphics}                 % Packages to allow inclusion of graphics
\usepackage{color}

\usepackage{graphicx}

\newcommand{\dnote}[1]{\textcolor{blue}{$\ll$\textsf{#1 --Dafna}$\gg$}}

\newcommand{\dpcomment}[1]{}

\usepackage{boxedminipage}
\usepackage{paralist}
\usepackage{tkz-graph}

\usepackage{graphics}                 % Packages to allow inclusion of graphics
\usepackage{color}                    % For creating coloured text
\DeclareGraphicsExtensions{.pdf,.gif,.jpg,.eps} % the formats we have images in

\makeatletter
\def\blfootnote{\xdef\@thefnmark{}\@footnotetext}
\makeatother

{\end{enumerate}}

\usepackage{tikz}
\usetikzlibrary{backgrounds}
\usepackage{tikz}
\usetikzlibrary{backgrounds}
\makeatletter

\tikzset{%
  fancy quotes/.style={
    text width=\fq@width pt,
    align=justify,
    inner sep=1em,
    anchor=north west,
    minimum width=\textwidth,
  },
  fancy quotes width/.initial={.8\textwidth},
  fancy quotes marks/.style={
    scale=8,
    text=white,
    inner sep=0pt,
  },
  fancy quotes opening/.style={
    fancy quotes marks,
  },
  fancy quotes closing/.style={
    fancy quotes marks,
  },
  fancy quotes background/.style={
    show background rectangle,
    inner frame xsep=0pt,
    background rectangle/.style={
      fill=gray!25,
      rounded corners,
    },
  }
}

\makeatother

%% file: 0-abstract.tex
\begin{abstract} 

A significant portion of web search queries directly refers to named entities. Search engines explore various ways to improve the user experience for such queries. We suggest augmenting search results with {\em trivia facts} about the searched entity. Trivia is widely played throughout the world, and was shown to increase users' engagement and retention.

Most random facts are not suitable for the trivia section. There is skill (and art) to curating good trivia. In this paper, we formalize a notion of \emph{trivia-worthiness} and propose an algorithm that automatically mines trivia facts from Wikipedia. We take advantage of Wikipedia's category structure, and rank an entity's categories by their trivia-quality. 
Our algorithm is capable of finding interesting facts, such as Obama's Grammy or Elvis' stint as a tank gunner. In user studies, our algorithm captures the intuitive notion of ``good trivia'' 45\% higher than prior work. Search-page tests show a 22\% decrease in bounce rates and a 12\% increase in dwell time, proving our facts hold users' attention.
\end{abstract}

% \dnote{and increase dwell time? CTR shows interest?}.

%From board games to mobile apps, newspapers and pubs, trivia is widely played throughout the world. From a business perspective, trivia was shown to increase customers' engagement and retention.

%Trivia is about unusual bits of everyday knowledge; however, most random facts are not suitable for the trivia section. There is skill (and art) to coming up with good trivia facts. In this paper, we formalize a notion of what is \emph{trivia-worthy} and propose an algorithm that automatically mines trivia facts from Wikipedia articles. We take advantage of the category structure of Wikipedia, and rank an entity's categories by their trivia-quality. 

%Our algorithm is capable of finding interesting facts, such as Obama's Grammy Award win, or Elvis' stint as a tank gunner. User studies demonstrate that our algorithm manages to capture the intuitive notion of ``good trivia''.

%% file: 1-intro.tex
\section{Introduction}
%(This isn't the right style file -- change later)

%\setlength{\epigraphwidth}{0.95\linewidth} 

%\begin{quote}
%\setlength{\epigraphwidth}{.98\linewidth}
%\setlength{\epigraphrule}{0pt}

\epigraph{
Libraries may be full of facts, but finding beautiful trivia in those dry, dusty stacks is like panning for gold. The glittering grains are few and far between. As the introduction to one early trivia book says, there is a difference between ``the flower of trivia and the weed of minutiae.'' Or, to put it another way, all trivia may be facts, but not all facts are capital-T Trivia. I can't spell out the difference, but I know it's there. ``Comedian Albert Brooks attended Carnegie Tech in Pittsburgh'' is a fact. So is ``Comedian Albert Brooks is five-foot-ten-inches tall'' -- not that interesting unless you're his tailor. But ``Comedian Albert Brooks had to change his name because he was born Albert Einstein''? Ah. That's trivia. 
}{Ken Jennings \cite{jennings2007brainiac}}

%\dnote{Ido -- it's already inside a citation. :)}

Today, anybody with a smartphone has access to far more information than even ``Jeopardy!'' champion Ken Jennings could ever recall. Despite this, trivia games are more popular than ever. From board games to mobile apps, newspapers and pubs, trivia is widely played throughout the world.

In this paper, we tackle the problem of {\bf automatically extracting trivia facts} from Wikipedia. 
This task, while seemingly lighthearted, has real-world applications. In particular, we are motivated by its application to \emph{search}.

%What does it mean to search? 
In many cases, search is a goal-driven activity: first, there is some information need, specific or general (e.g., the current time in Alofi, or alternatively, a cure for Zika). We then approach a seemingly-omniscient mechanism, in the form of a search engine, to answer the need. We wait, get the answer, and then go on --- presumably towards some other well-defined information need. This is the prevailing view in the information-retrieval community, which places the search-engine and the user at opposing roles: one is the producer of search topics, while the other is the consumer, and acts merely as a passive librarian, looking up the facts. However, many users do not share this utilitarian view~\cite{andre2009}. For them (or for some of their queries), search is an exploratory activity, and at some stages of the information-gathering process, less-than-relevant results are welcome~\cite{spink1998}.

With this in mind, we note that a significant portion --- over 50\% --- of web search queries directly refers to named entities~\cite{mika13entity,Yin2010}. Modern web search engines explore various technologies to improve user experience for these types of queries. Such technologies include clustering of search results for disambiguation, related entity recommendation, and the presentation of rich ``entity cards'', which include key aspects of the entity, directly on the search engine results page (SERP)~\cite{Bota16playing}. 
Recently, Miliaraki et al.\ \shortcite{miliaraki2015} demonstrated a search system which, in addition to the search results, surfaced entities related --- but not necessarily directly --- to the query. 
%
%The affiliated query was related to the search query by similarity defined over a large knowledge base. For example, a search for Selena Gomez could trigger a reference to Kevin James, who co-starred with her in a movie. 
%
%Other measures of similarity could be co-occurrence in text, or geographical or cultural proximity (for example, of cities).
Importantly, surfacing other entities was proven to be an effective vehicle for drawing searchers to an exploratory activity, 
%: the system did entice users to explore, 
thereby increasing engagement. 
%This was examined along several dimensions. In essence, the highest engagement was registered when the mentioned entity was a person (as compared to location, or a movie). Additionally, search terms that contain just the entity (without modifiers) are correlated with high related-entity engagement, when compared to queries that follow the entity with the ``news'' or ``photo'' modifiers (while the ``movie'' modifier was also shown to boost engagement). Finally, there is a strong match between the age of the person suggested as a linked entity, and the age of the searcher; the import is that whatever one considers to be a worthwhile distraction is highly personalized.

We propose augmenting search results with \textbf{trivia facts} that are related to the searched entity. We believe trivia facts could contribute to the user experience around entity searches; even a small impact on this type of queries can translate into a significantly improved user experience. 
%\ig{maybe refer to that news article about the guy who made fortune by presenting trivia?}

There are multiple reasons to believe that trivia can indeed contribute to the user experience.
Business case studies \cite{TriviaCase} have shown that trivia helps increase user engagement and revenue. A man who tweets random facts has over 18 million followers, and makes about \$500,000 a year from sponsored links~\cite{UberFacts}. 
%And despite the fact that anybody with a smart-phone can access far more information than even ``Jeopardy!'' champion Ken Jennings could ever remember~\cite{ferrucci2010building}, bar trivia games and trivia apps are more popular than ever. 
%, which demand that people put away their devices to test their analog library of knowledge, 

%However, there is skill (and art) to coming up with good trivia facts. 
However, there is skill (and art) to coming up with good trivia. 
%Coming up with good trivia facts is difficult. 
In a recent experiment, professional trivia curators managed to find trivia facts for merely ten entities per day, on average \cite{prakash2015did}.
The process is expensive and hard to scale.

In order to automate the process of finding good trivia, one needs to characterize the notion of what is \emph{trivia-worthy}. In this work, we introduce and formalize two criteria that characterize good trivia: \emph{surprise} and \emph{cohesiveness}. 
Using our formulation, we propose an algorithm that automatically extracts trivia facts from Wikipedia articles.  
We take advantage of the category structure of Wikipedia, and rank an entity's categories by their trivia-quality. Our algorithm is capable of finding interesting facts, such as Obama's Grammy Award win, or Elvis' stint as a tank gunner. 

User studies with crowd-sourced workers show that our algorithm produces facts that capture the intuitive notion of ``good trivia''. In another study, we bought ads on search pages and measured the level of interest in trivia in a real life scenario. Trivia facts were generally found to arouse interest, while better trivia facts attract page views with lower bounce rates and longer dwell time. 
%\dnote{finishing sentence}

%\ig{If time allows, put more emphasis on the problem of extracting good trivia, the definition of trivia, the idea of using the Wikipedia categories, and the evaluation against the baseline.}

%Our contributions are:
%\begin{compactitem}
%\item We formalize a notion of \emph{trivia-worthy} mathematically.
%\item We implement an algorithm 
%\end{compactitem}

%\end{quote}

%% file: 2-problem.tex
\section{Problem Formulation}
\label{sec:problem}
\remove{
\begin{itemize}
\item What are we looking for?
\item Running example: categories of Hedy?
\item Surprise. How to formulate
\item Running example: what categories are surprising
\item Cohesiveness
\item Running example: Surprising and cohessive
\item Anything else? Page views?
\end{itemize}
}

\begin{figure}[t]
\centering
\includegraphics[width=0.95\linewidth]{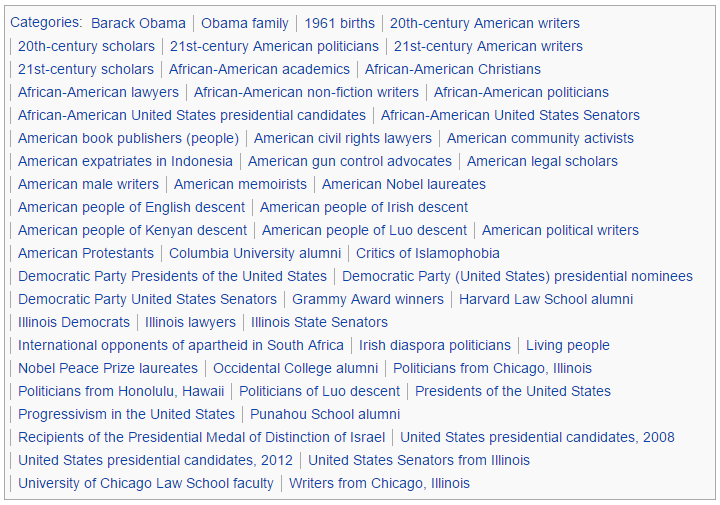}
\caption{\small Categories from Obama's Wikipedia page %\dnote{I got it back -- I think the point here is to show there are a lot, not to read them} 
\label{fig:ocats}}
\end{figure}

Our goal is to automatically find trivia facts about entities.  We first consider possible sources of such facts. Wikipedia is a natural choice for this purpose because of its wide coverage. However, Wikipedia articles are written in natural language; working with short textual units (e.g., sentences), one must deal with anaphora resolution, long-range references, and other context-related problems.  
Thus, we focus on Wikipedia's \emph{category structure}. Categories are sets of articles with a shared topic, such as ``History of France'', ``Philosophy of mind'', or ``Biological concepts''. An article can belong to multiple categories. For example, Barack Obama's categories include ``Presidents of the United States'', ``Columbia University alumni'', and ``Grammy Award winners'' (see Figure \ref{fig:ocats}). 
Importantly for us, categories are cleaner than sentences, while often capturing the most interesting aspects of the article \cite{chernov2006extracting,nastase2008decoding}.
%\ig{We should probably cite an article or two here, such as: Extracting Semantics Relationships between Wikipedia Categories by Chenov et al.\ or Decoding Wikipedia Categories for Knowledge Acquisition by Nastase}

Given an article, we want our algorithm to rank its categories, such that the top-ranked categories should be most suitable for the trivia section. Therefore, we need to formalize the notion of \emph{trivia-worthy}. The Merriam-Webster dictionary defines trivia as
%
%\begin{displayquote}
{\it ``Unimportant facts or details. Facts about people, events, etc., that are not well-known.''}
%\end{displayquote}
%
%\begin{compactitem}
%\item Unimportant facts or details
%\item Facts about people, events, etc., that are not well-known
%\end{compactitem}
%

One possible direction for detecting a trivia-worthy category would be to choose the category with the smallest number of articles. Presumably, a small category indicates a rare and unique property of an entity, and would be an interesting trivia fact. However, testing this path has shown it focuses on properties that were too narrow, in several senses: Most often, the smallest category focuses on a very specific identity aspect, usually obscure and uninteresting - ``Muhammad Ali is an alumni of Central High School in Louisville, Kentucky'' is not a good trivia fact - the specific high school has no importance to the reader and does not reflect on Ali's character. In other cases, when the entity belonged to a well-known family, band or group, the smallest category captured a well-known aspect of the entity, for example ``Michael Jackson was a member of the The Jackson 5''.

Indeed, trivia facts are often centered around uncommon knowledge. In other words, trivia facts are \emph{surprising}. For example, everybody who knows Obama probably knows he belongs to the ``Presidents of the United States'' category, but many people would be surprised to learn that he won a Grammy. On the other hand, we want facts that are also interesting and not obscure.

We begin by formulating our first property - \emph{surprise}.

%\ig{I find the notion of ``interestingness'' a bit confusing, especially since it subsumes ``surprising''. I think we might get criticism for using it. Seems like it should be something like ``good trivia'' or ``fun fact'', don't have a perfect idea in mind yet... }

\subsection{Surprise}
%\dnote{I changed the order back because it's important to give people the high level idea of why we're doing this, before we expect them to dive into formulas. Motivation! :) }

Surprise measures how unusual it is for a given article to belong to a category. In other words, we would need to define a similarity metric between an article $\article$ and a category $\catg$. Since a category is a collection of articles, our main building block will be a similarity metric between articles.
We denote article-article similarity by $\artsim(\article,\article')$, and defer its exact implementation details to Section \ref{sec:alg}.

%\dnote{we need a similarity metric -- defer it to the next section}

%\subsubsection{Surprise as Distance from Category}

Next, we extend the similarity between articles to similarity between articles and \emph{categories}.
A category $\catg$ is a set of articles.
%
%We model each category $\catg$ as a complete weighted graph \(G_\catg=(V_\catg,E_\catg)\). Each article in the category is represented by a vertex. Between every two vertices there is \dnote{don't use there's, isn't, etc. in a paper} a weighted edge indicating the respective article similarity. 
%
We define the similarity of an article $\article$ to category $\catg$ as the average similarity between $\article$ and the articles of $\catg$:
\[\artcatsim(\article,\catg)=\frac{1}{|\catg|-1}\sum\limits _{ \article\neq \article' \in \catg} \artsim(\article,\article')\]

An article is surprising w.r.t.~a category if its average similarity to the other articles is low. Thus, we define surprise as the inverse of the average similarity:
\[ \surprise(\article,\catg) = \frac{1}{\artcatsim(\article,\catg)} \]

%\dnote{note what I did here. We want high surprise to actually mean something is surprising. :)  This changes all the graphs, too.}
%\dmtsurel{Good idea.}

For example, consider 1940s Hollywood film actress Hedy Lamarr. When ranking her Wikipedia page categories by surprise, the top 5 results (out of 18) are, in order (top is most surprising): %\dpnote{Which are least surprising?}

\begin{mdframed}[roundcorner=10pt]
	\centering
``20th-century Austrian people'' \\ ``Women in technology'' \\ ``Radio pioneers'' \\ ``American anti-fascists'' \\ ``American people of Hungarian-Jewish descent'' 
\end{mdframed}
\newpage

\subsection{Cohesiveness}

Looking at Hedy Lamarr's most surprising categories, some of them do not fit our intuitive idea of trivia. For example, regarding her Austro-Hungarian descent: Wikipedia chose to list it because of its charter to record the minutiae of famous people's biographies, but the detail itself is not particularly trivia-worthy. %\dmtsurel{It actually is a pretty trivia-worthy aspect. Being a Jew in 1930s Austria, her husband's business ties with Mussolini, rescuing her mother from the Nazis. So we shouldn't be too harsh with this fact.}
%\dnote{David -- if it said ``her husband's business ties with Mussolini'' or ``she rescued her mom from Nazis'' that would be awesome trivia, but it doesn't}

However, other aspects of her life are less mundane. For example, this Hollywood star had also invented radio encryption (more precisely, she had patents in frequency-hopping, spread-spectrum technology\footnote{As well as inventions in traffic lights and soft drinks. Did you know {\em that}?}). Yet, our notion of surprise ranks the fact similarly to how it ranks her (less exciting) Austrian pedigree. 

Note that all five categories are, indeed, surprising; many people who know Hedy Lamarr as a movie actress are probably not aware of these aspects of her life. Therefore, we conclude that surprise is not enough.

However, the thing that makes the group ``Radio pioneers''  more suitable for our purposes is somewhat harder to define. 
Intuitively, being of Hungarian-Jewish descent seems more arbitrary than being a radio pioneer; being Austrian says less about the person than being a woman in technology. We believe that elusive notion is related to the \emph{cohesiveness} of the group: People born in Austria can come from all walks of life. Look at the Wikipedia category ``20th-century Austrian people'', and you will find lawyers, architects, painters, physicians, and World War I diplomats. On the other hand, radio pioneers seem like a more close-knit group.
%, sharing not just knowledge of a particular technology, but also personal choices throughout their life, enforcing their bias towards this specific area of interest.

Thus, we define our second metric, \emph{cohesiveness}. We define cohesiveness of category $\catg$ as the average similarity between pairs of articles from $\catg$:
\[\cohesive(\catg)=\frac{1}{{|\catg| \choose 2}}\sum\limits _{\article \neq \article'}\artsim(\article, \article')\]

When ranking Hedy Lamarr's categories by cohesiveness, the top-5 categories are, in order:  

\begin{mdframed}[roundcorner=10pt]
	\centering
``Metro-Goldwyn-Mayer contract players'' \\ ``Actresses from Vienna'' \\ ``Austrian film actresses'' \\ ``20th-century Austrian actresses'' \\ ``American film actresses''
\end{mdframed}

These categories are indeed cohesive, in the sense that they are not arbitrary details. 
%they are more specific: here it's not enough to be just Austrian; on top of that, you also have to be in the movie industry. In general, the groups here are more specific than the ones in the list of surprising categories, above: cities instead of countries, specific employers instead of industry sectors, etc. 
%
However, to be fair, this list is not yet a good set of trivia facts. For that, we have to consider both surprise and cohesiveness, which we do below.

\subsection{Tying it Together}

%\dnote{about why ratio has a natural interpretation -- how much above expectation you are}

%\dnote{Running example:}

\begin{figure}[t]
%\centering
\includegraphics[width=\linewidth]{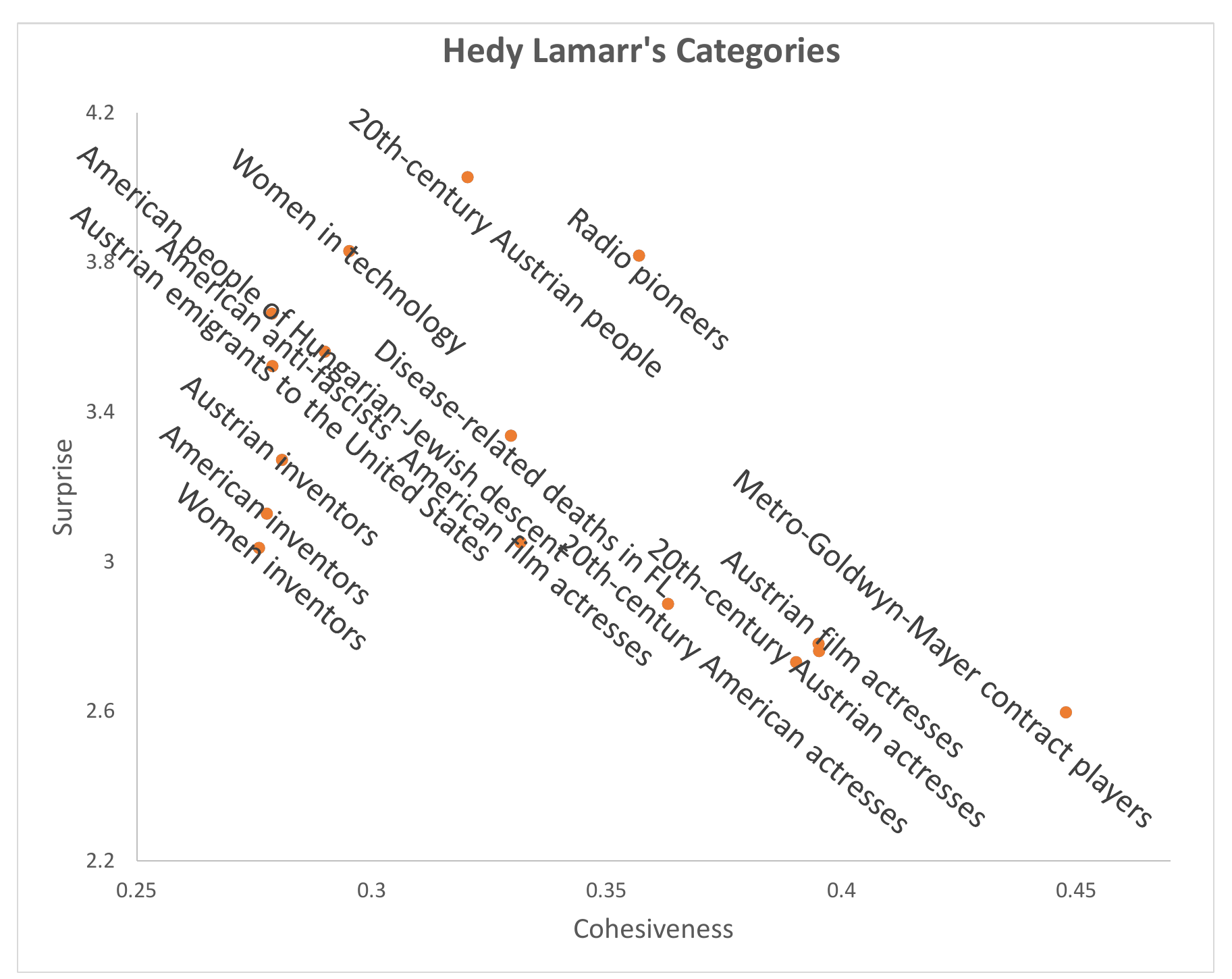}
\caption{\small Cohesiveness plotted against surprise for the categories of Hedy Lamarr. To reduce label clutter, we omit some of the labels. \label{fig:HedySCgraph}}
%\dmtsurel{better?}
%\dpnote{Some dots are now orphaned. And the axis labels are in a tiny font}\dnote{These dots are fine -- just write in the caption that ``for the sake of presentation, we omit some labels'' or something along those lines. Axis is indeed too small}\dmtsurel{Did another iteration.}
\end{figure}

\begin{figure*}[t]
\centering
\includegraphics[width=1\linewidth]{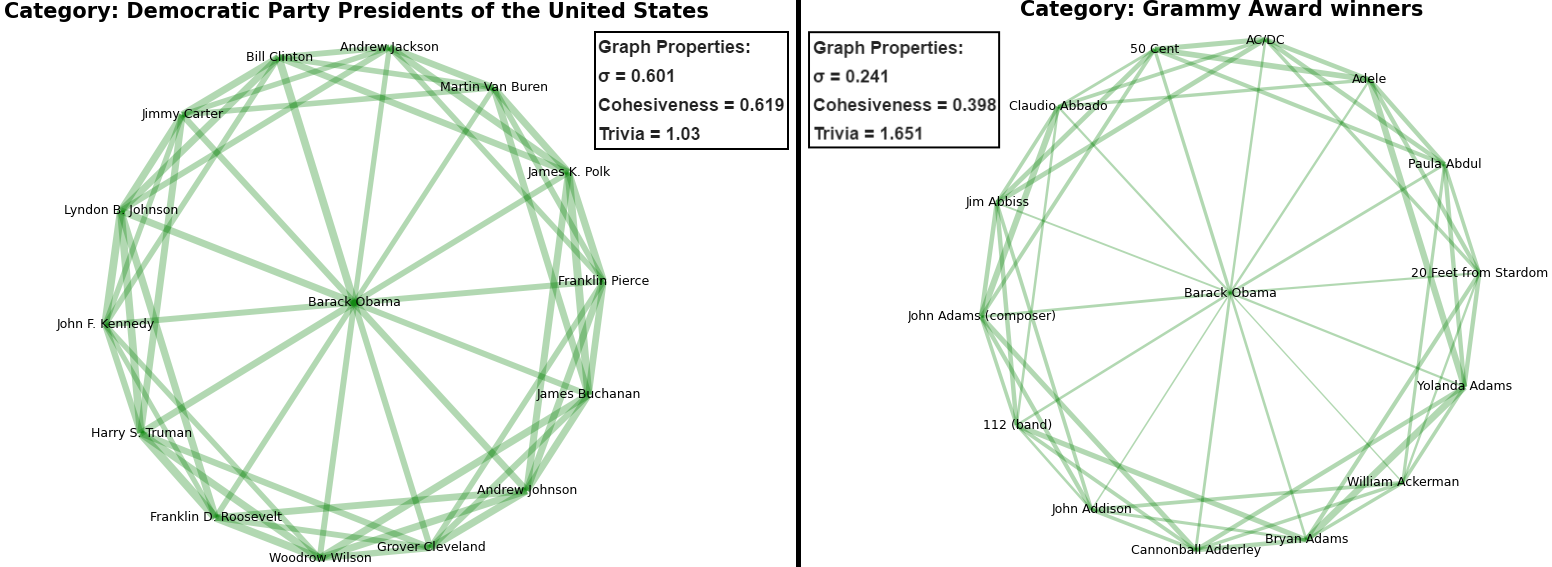}
\caption{Similarity graphs for two categories containing Barack Obama. Thicker edges are more similar. For visualization reasons, not all nodes and edges are shown. \label{fig:obama_category_graphs} }
\end{figure*}
%\dmtsurel{Where does figure \ref{fig:obama_category_graphs} go now? Is it big enough?}

%\dnote{formulation}
We have just defined two properties -- surprise and cohesiveness. We now wish to combine them into a notion of \emph{trivia-worthiness}. 

Figure \ref{fig:HedySCgraph} shows the categories of Hedy Lamarr. The x-axis represents cohesiveness, and the y-axis represents surprise. Intuitively, a category is trivia-worthy if it is high on both surprise and cohesiveness scores. Therefore, we would like to define a score that is monotonic in both dimensions.

We define the trivia-worthiness of a category $\catg$ w.r.t. article $\article$ as the product of cohesiveness and surprise. 
\[\interest(\article,\catg)={\cohesive(\catg)}\cdot{\surprise(\article,\catg)} \]

%\dpnote{It's rather arbitrary that we combine the two values with multiplication. Strictly speaking, given two dimensions, we want to combine them with a function that is at least monotonic in each, and beyond that there are many options (cue social choice work). I could think of alternatives to multiplication as the combiner, and we should probably expect some reviewer sentiment here.}

%\dmtsurel{I think the next paragraph explains why it's not arbitrary.}

%\dmtsurel{But Dan is right that we can rearrange these paragraphs to make it clear from the start that it isn't arbitrary.}

There are many other ways to combine the two scores. However, multiplication has a natural interpretation.
Note that surprise is defined as the inverse of the average similarity of $\article$ to the category. Thus, the multiplicative formula measures how similar the article is to the category, compared to the average similarity of articles from the same category. In other words, we measure whether the article is more similar or less similar to the category than was expected.
\[\interest(\article,\catg)= 
\frac{\cohesive(\catg)}{\artcatsim(\article,\catg)} \]

A value of $\interest$ around one means, by definition, that the average distance between $\article$ and $\catg$ is similar to the cohesiveness of $\catg$, which indicates that the article is typical for that category: it is similar to other articles in the category just as much as the average article. 

A value of $\interest$ much lower than one indicates that the article is more similar to other articles than the average, and could be an {\em exemplar}. It is a prominent member of the category, and would not be good trivia.

Now, a value of $\interest$ higher than one means that the article in question is not so similar to the category. In some sense, it is an {\em outsider}, which might make good trivia.

\xhdr{Example}
Figure \ref{fig:obama_category_graphs} illustrates our ideas. The figure shows similarity among articles from the 
``Democratic Party Presidents of the United States'' category (left), compared to articles from ``Grammy Award winners'' (right). Each node is an article, and the edge weight represents similarity. Obama is in the center of both graphs.

First, we look at the edges from Obama to other articles. Being a Grammy winner is much more \emph{surprising}, as demonstrated by the thinner edges between Obama's node and the rest of the graph --- a similarity score of only 0.241, compared to 0.601 for the democratic presidents.

Next, we move on to \emph{cohesiveness}.
The cohesiveness of the two different categories can be seen in Figure \ref{fig:obama_category_graphs} as the average thickness of the edges. The Grammy winners have a cohesiveness score of 0.398, as they are mostly well known musicians, although from different genres. Presidents, on the other hand, are all leading US politicians, and are a well-knit group, earning them a score of 0.619.
%\dmtsurel{I refined some of the details in the explanation. 0.398 is not such a low score (compare Hedy graph).}

Finally, we look at Obama's edges, compared to the rest of the edges in the category.
We can see that in the democratic presidents graph, Obama has high similarity to the other nodes, but so is the average similarity in the graph. The \(\interest\) value is \(\frac{\cohesive(\catg)}{\artcatsim(\article,\catg)}=\frac{0.619}{0.601}=1.030 \), which indicates that Obama is a typical article in the category. In the Grammy category, however, the similarity between Obama and the other nodes is much weaker than the average similarity in the category. The \(\interest\) value is \(\frac{0.398}{0.241}=1.651 \), which indicates that Obama is not a typical Grammy winner.

%The cohesiveness of two different categories is shown in figure \ref{fig:obama_category_graphs} as the average thickness between all edges in the graph. We can see that the Grammy category is less cohesive, since its average edge width is thinner.

%% file: 3-algorithm.tex
\section{Algorithm}
\label{sec:alg}

For a formal description of our method, see Algorithm~\ref{alg:top}. $K$ is a parameter, described below. In each category, we compute 
cohesiveness and surprise with regard to the input article, and combine them into a trivia score. 

An important component in our formulation was an article-article similarity metric, $\artsim(\article,\article')$ (computed by the function {\sc ArticleSimilarity}). We now discuss its implementation details. 

%\newpage %avoiding break

\subsection{Article Similarity} 

When choosing a metric for article comparison, standard methods such as cosine similarity between term frequency vectors proved to be inadequate.
For example, when comparing the articles ``Apple'' and ``Orange'', cosine similarity was only $0.026$, even though both are fruit. ``Apple'' and ``Barack Obama'' had a higher similarity, $0.059$. There are two main problems underlying the usual similarity methods:

\begin{algorithm}[h!]
\caption{Top Trivia algorithm}\label{alg:top}
\begin{algorithmic}
\Function{TopTrivia}{$inputArticle$}
\For{\text{every category} $C$ \text{of $inputArticle$}}
\State $surprise\gets \Call{Surprise}{inputArticle, C} $
\State $cohesiveness\gets \Call{Cohesiveness}{C} $
\State $C.trivia\gets cohesiveness * surprise $
\EndFor
\State \textbf{return} \text{category $C$ with maximum $trivia$ score}\
\EndFunction
\Statex

\Function{Surprise}{$inputArticle, category$}
\State $sum, count\gets 0 $
\For{\text{every article} $a\neq inputArticle$\text{ in category} $C$}
\State $similarity\gets \Call{ArticleSimilarity}{inputArticle, a} $
\State $sum\gets sum+similarity $
\State $count\gets count+1 $
\EndFor
\State $similarityToCategory\gets sum/count $
\State $surprise\gets 1/similarityToCategory $
\State \textbf{return} $ surprise $\
\EndFunction
\Statex

\Function{Cohesiveness}{$category$}
\State $sum, count\gets 0 $
\For{\text{every pair of articles} $a1 \neq a2$ \text{in category} $C$}
\State $similarity\gets \Call{ArticleSimilarity}{a1, a2} $
\State $sum\gets sum+similarity $
\State $count\gets count+1 $
\EndFor
\State $cohesiveness\gets sum/count $
\State \textbf{return} $ cohesiveness $\
\EndFunction
\Statex

\Function{ArticleSimilarity}{article1, article2}
\State $K\gets 10$
\State $T1\gets \Call{TopTFIDF}{article1,K} $
\State $T2\gets \Call{TopTFIDF}{article2,K} $
\State $similarity \gets\artsim(article1,article2) \text{ using equation \ref{eq:1}} $
\State \textbf{return} $ similarity $\
\EndFunction

\end{algorithmic}
\end{algorithm}

\begin{compactitem} 
\item  We are looking for relatively {\bf broad similarity} (for example, ``both people sing in rock bands''); we do not necessarily need details to be similar. Long articles, in particular, can add a significant amount of noise. 
\item Term frequency vectors look for exact matches between terms, so many {\bf semantic similarities} are lost (even after stemming and normalization). 
\end{compactitem}

%To address these problems, our method for measuring article similarity used word2vec, whose context window captures semantic similarities far beyond simple lexical comparisons. Instead of measuring all terms in an article, our algorithm removed the noise by using only the top \(K\) tf-idf terms. The algorithm optimized the parameter \(K\) for differentiating pairs of similar articles from pairs of articles that are not. \(K=10\) was found to be optimal. 

To address the first problem ({\bf broad similarity}), we do not use all words in an article. Instead, we compute TF-IDF scores for all words in the documents. TF-IDF measures how important a word is in a document, given a corpus. For our corpus, we used a sample of $10,000$ articles from the English Wikipedia. We used standard text normalization techniques such as stemming, stopword removal and case folding. We removed terms appearing in less than 10 documents.

We restrict ourselves to the top \(K\) TF-IDF terms of each article. 
In our experiments we used \(K=10\), after testing showed it balanced between noise reduction and retaining important information. %\dmtsurel{I added this explanation for the K parameter}
For example, table \ref{TopTFIDF} displays the top 10 TF-IDF terms for the articles ``Sherlock Holmes'' and ``Dr. Watson'', the main characters in the detective stories of Arthur Conan Doyle, and the article ``Hercule Poirot'', another fictional detective from stories by Agatha Christie.

While these words seem to capture the gist of the three characters, there are almost no exact matches. Holmes and Poirot, for example, share exactly one word -- detective. There are, however, many semantically related words, such as murder/detective, novels/adventures. 

To address the problem of {\bf semantic similarity}, we compute word similarity $\wordsim(w_1,w_2)$ using the word2vec representation \cite{mikolov2013efficient}. We used a pre-trained model, trained on a Google News corpus of about 100 billion tokens using a neural network to produce a 300-dimensional vector space of word embeddings. 
%Each word is represented in this pre-trained model by an embedded vector, and word similarity is then defined to be the cosine similarity between the corresponding vectors.
%\[\wordsim(w_1,w_2)=\frac{\sum\limits _{i=1}^{n}word1_{i}word2_{i}}{\sqrt{\sum\limits _{i=1}^{n}word1_{i}^{2}}\sqrt{\sum\limits _{i=1}^{n}word2_{i}^{2}}}\]
Word similarity is in the range [-1,1], where higher values indicate stronger similarity. 
Interestingly, word2vec is known to capture semantic similarities \cite{baroni2014don, Kenter:2015:STS:2806416.2806475}. 
For example: 
\begin{align}
\wordsim(``christie",``doyle") & = 0.529 \nonumber \\
\wordsim(``novels",``adventures") & =  0.423 \nonumber \\
\wordsim(``curtain",``scarlet") & =  0.163 \nonumber
\end{align}

\begin{table}[t]
\centering
\small
\caption{\small Top TF-IDF Terms}
\label{TopTFIDF}
\begin{tabular}{@{}lll@{}}
\toprule
 Sherlock Holmes & Dr. Watson & Hercule Poirot \\ \midrule
 holmes          & watson     & murder         \\
 sherlock        & holmes     & christie       \\
 watson          & sherlock   & hastings       \\
 adventures      & adventures & detective      \\
detective       & doyle      & novels         \\
 conan           & portrayed  & curtain        \\
 doyle           & conan      & belgian        \\
 stories         & stories    & solve          \\
 scarlet         & detective  & mystery        \\
 bohemia         & doctor     & adaptation     \\ \bottomrule
\end{tabular}
\end{table}

%\dnote{does your word2vec version actually have negative numbers? Did we ever see them?}

%\dmtsurel{It's possible to have negative values, but since we take the maximum over the set it is unlikely. }

%\xhdr{Article Similarity}
To compute similarity between articles, we look at the similarities of their top words.
Let $T_1$ and $T_2$ be the sets of top-$K$ TF-IDF terms for two articles, $\article_1$ and $\article_2$. For each TF-IDF term in $T_1$, we find the most similar term in $T_2$ (and vice versa, to keep the definition symmetric) and sum up these similarities. We use a weighted formula to give more weight to terms with higher TF-IDF scores and normalize the result to the range [-1,1]:
\begin{align}
%\begin{equation} 
%\begin{split}
\artsim(\article_1,\article_2)=  &  \label{eq:1} \\
\frac{1}{\mathcal{Z}}  \sum\limits _{i=1}^{K} w(i)\cdot &
\ensuremath{(\max\limits _{1\leq j\leq K}}\wordsim(T_1[i],T_2[j])+  \ensuremath{\max\limits _{1\leq j\leq K}}\wordsim(T_2[i],T_1[j]) \ ) \nonumber
%\end{split}
%\end{equation}
\end{align}

We experimented with several weighting schemes, and chose a linear one: $w(i)= K-i+1$, with normalization factor $\mathcal{Z}= {2\cdot\binom{K+1}{2}}$.

When comparing articles using this method, the articles ``Apple'' and ``Orange'' had a similarity score of $0.3$, compared to only $0.11$ for ``Apple'' and ``Barack Obama''. ``Sherlock Holmes'' and ``Hercule Poirot'' had a similarity score of $0.513$.

\subsection{Practical Considerations}
To improve efficiency and allow reuse of intermediate results, we used \emph{caching} throughout our algorithm, writing values to both memory and file system. To increase speed, we \emph{parallelized} the algorithm so it could (1) process several articles simultaneously, and (2) rate the trivia-worthiness of several categories simultaneously for each article. 
%\dnote{David -- do you have some running-time numbers here?} \dmtsurel{It's not easy to measure. Tell me if it's absolutely critical and I'll work on it.}

When computing surprise and cohesiveness for large categories, one needs to compute similarity between $O(n^2)$ pairs of articles. To speed the computation up, we randomly sample a subset of the articles instead, and computed similarity between all pairs in the subset. In our experiments, we found that 50 articles are usually enough to obtain results that are very close to the results of using the full set of articles.

%% file: 4-evaluation.tex
\section{Evaluation}
\label{sec:eval}

In this section, we evaluate our algorithm empirically.
We have compared the following algorithms:
\begin{compactitem}
\item {\bf Wikipedia Trivia Miner (WTM)} \cite{prakash2015did}: A ranking algorithm over Wikipedia sentences, which learns the notion of interestingness using domain-independent linguistic and entity based
features. The supervised ranking model is trained on existing user-generated trivia data available on the Web.
\item {\bf Top Trivia}: The highest ranking category in our algorithm ranking.
\item {\bf Middle-ranked Trivia}: Using middle-of-the-pack ranked categories, as ranked by our algorithm.
\item {\bf Bottom Trivia}: Using the lowest-ranked categories by our algorithm. 
\end{compactitem}

%\xhdr{Data}
We collected article and category data for our experiments via the Wikipedia web API using the Pywikibot framework~\cite{Pywikibot} and an adapted version of the Wiki2Plain interface~\cite{Wiki2Plain}. 
%We used standard text normalization techniques including stemming, stop word removal, and case folding.  

We created a dataset of 400 popular Wikipedia articles about people, based on a list of the most viewed pages over the week of July 10-16, 2016 \cite{PopularPages}. The list contains a diverse range of popular people, including politicians, sportspeople, scientists, actors, writers, singers, historical figures and other people of interest. 

However, the popularity of a page does not necessarily indicate that it contains good trivia. To ensure a fair comparison, we restricted ourselves to pages where both our algorithm and WTM found good trivia. 
In particular, we selected articles for which
the trivia fact had a score in the top 50\% of facts in both our algorithm and the WTM rankings. This resulted in a dataset of trivia facts for 109 articles. 

For every article, we produced the single trivia fact for each of the algorithms. The textual format for our facts is ``$a$ is in the group $\catg$''. Table~\ref{obamaFacts} shows an example of the trivia facts produced for the article ``Barack Obama''. Our data and code are available at \url{https://github.com/DMTsurel/FunFacts}
%\ig{very nice example, I like it a lot}

\begin{table}[t]
\centering
\small
\caption{\small Top fact returned by each algorithm for the article ``Barack Obama''}
\label{obamaFacts}
\begin{tabular}{p{1.5cm}|p{5.5cm}}
\toprule
Algorithm & Fact \\ \midrule
Top & Barack Obama is in the group of Grammy Award winners \\ \midrule
Middle & Barack Obama is in the group of African-American lawyers \\ \midrule
Bottom & Barack Obama is in the group of Obama family \\ \midrule
WTM & Besides his native English, Obama speaks some basic Indonesian, having learned the language during his four childhood years in Jakarta. \\ \bottomrule
\end{tabular}
\end{table}

\subsection{Trivia Evaluation Study} 

Evaluation of trivia facts is a subjective matter. Therefore, a key part of our evaluation is based on a user study we performed using crowd-sourced work.

For each of the 109 articles, we computed four trivia facts: our top, middle and bottom facts, as well as WTM. Each fact was presented to five crowd workers, for a total of 2180 evaluations. To increase reliability, we restricted  workers' location to the US, and their approval rate to above 80\%. We made our task available to \emph{Mechanical Turk Masters} only (a qualification given by Mechanical Turk to workers who perform consistently well across a wide range of tasks).  %Masters must continue to pass our statistical monitoring to maintain the Mechanical Turk Masters Qualification.
%\ig{can we add a bot more details here? can we say something about the quality of the turkers? did we use ``trap'' questions? anything to give more sense of reliability to the reader}

The workers were presented with the fact and asked to express their level of agreement with the following statements:

\begin{compactitem} 
\item Trivia-worthiness: \textit{``This is a good trivia fact''. }
\item Surprise: \textit{``This fact is surprising''. }
\item Personal knowledge: \textit{``I knew this fact before reading it here''.}
\end{compactitem} 

Workers could agree or disagree with each statement, or reply that they could not understand the fact. For each statement, the \(majority\) \(opinion\) of a fact is the answer agreed on by at least 50\% of workers. Five evaluations had missing answers for the trivia-worthiness statement. In two of these, no majority was reached because of the missing answer, so they were removed from the results.

\begin{figure}[b]
\centering
\includegraphics[width=0.98\linewidth]{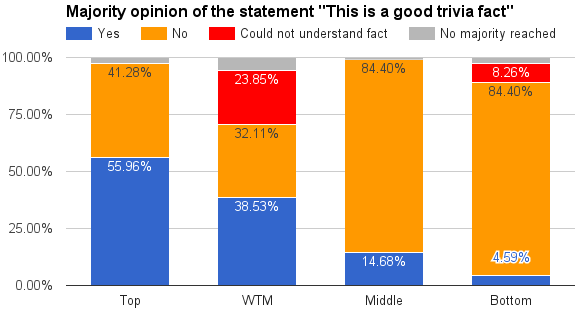}
\caption{\small Majority opinion about facts being trivia-worthy, by algorithm \label{fig:MTurk_good_trivia}}
\end{figure}

\xhdr{Results}
Figure \ref{fig:MTurk_good_trivia} shows the percentage of facts that a majority of users ranked as trivia-worthy, by algorithm. As expected, the facts ranked ``top'' by our algorithm outperform the ``middle'', which outperform ``bottom''. Our algorithm proved to be significantly better at finding trivia-worthy facts than the WTM baseline: 56\%, compared to only 38.5\% for WTM (\(p < 0.01 \), Pearson's chi-squared test).

When looking at the type of majority (3, 4 or 5 users), we notice that $32.8\%$ of the facts ranked as trivia-worthy by our algorithm achieved a perfect agreement (5 users), compared to only $11.9\%$ for WTM.
%\ig{can we also make a case for WTM having the highest portion of disagreement (no majority reached) as an indicator of confusion, or are the numbers too small?}

%\ig{which statistical test was used?}
%\dmtsurel{Added. Tell me if I should use a different test.}

Facts that could not be understood are the worst type of facts -- not only are users presented with facts that are not good trivia, but they leave the users confused. Our algorithm had no such facts, compared to 23.9\% for the WTM baseline and 8.3\% for our bottom baseline. This is an advantage of using categories for facts, as they are self-contained pieces of information. The WTM baseline used sentences from the Wikipedia text, which are sometimes left out of context even after trying to remove such sentences using co-reference resolution~\cite{prakash2015did}. For example, the WTM baseline fact for Leonardo DiCaprio was ``The project achieved a worldwide box office take of \$147 million.'' Users did not know what project was referenced in this sentence, so they could not understand the fact. Our bottom baseline also had several confusing facts. For example, the ``William Shakespeare'' category was ranked worst for the article William Shakespeare, and users were confused by the fact ``William Shakespeare is in the group of William Shakespeare''.
%\ig{can we also make a case for WTM having the highest portion of disagreement (no majority reached) as an indicator of confusion, or are the numbers too small?}
%\dmtsurel{Too small, only about 6 cases.}

We examined instances where our algorithm failed while WTM managed to find an interesting fact. Our algorithm examines only facts formulated as categories, so it will miss anecdotes that do not pertain to a set of articles. For example, most workers did not think the fact ``Beyonce is in the group of Shoe designers'', found by our algorithm, was trivia worthy. The trivia fact suggested by WTM was ranked as good trivia: ``On January 7, 2012, Beyonce gave birth to her first child, a daughter, Blue Ivy Carter, at Lenox Hill Hospital in New York''. The latter fact is too specific to be captured by a category.

\begin{figure}[h]
\centering
\includegraphics[width=0.98\linewidth]{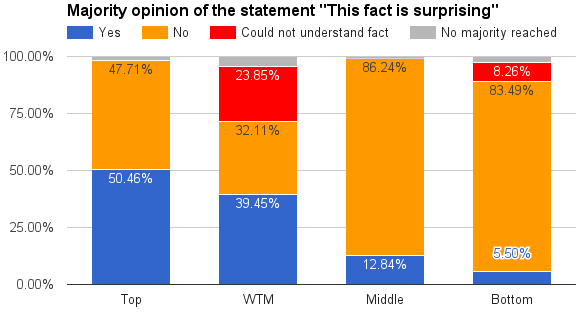}
\caption{\small Majority opinion about facts being surprising, by algorithm \label{fig:MTurk_surprise}}
%\dpnote{Do we have some more data on the type of majority? Is it 100/0, 80/20, 60/40?}
\end{figure}

Results for surprise were similar (Figure \ref{fig:MTurk_surprise}). Facts ranked as the top category were more surprising to users than those in the middle and bottom. 50.5\% of our algorithm's top results were surprising to users, 47.7\% were not surprising, and there were 0\% where users could not understand the fact. The WTM algorithm had 39.5\% of its facts ranked as surprising, 32.1\% as not surprising, and 23.8\% could not be understood (\(p < 0.01 \), Pearson's chi-squared test).
%\dnote{elaborate a little -- bad form to show a figure without explaining it} \dmtsurel{done}

\begin{figure}[h]
\centering
\includegraphics[width=0.98\linewidth]{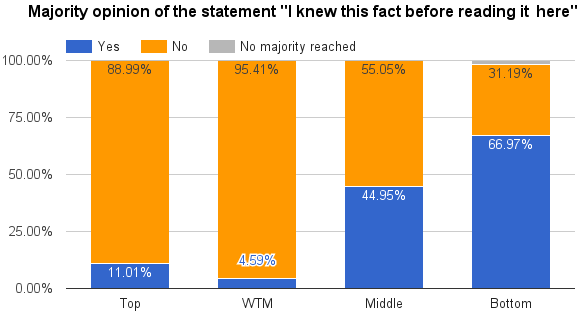}
\caption{\small Majority opinion about personal knowledge of facts, by algorithm \label{fig:MTurk_knew}}
\end{figure}

Figure \ref{fig:MTurk_knew} shows the percentage of facts that a majority of users knew previously. Almost all facts in both our algorithm's top choice and WTM were previously unknown to users.
However, when choosing the middle or bottom categories, the likelihood of being familiar with the facts is much higher. 
This indicates that our ranking method works well in terms of filtering out well-known facts. 

To test our hypothesis that good trivia is based on the element of surprise, we consider the contingency table of trivia-worthiness and surprise  (Table~\ref{triviaSurprise}).
We see that there is indeed strong correlation between surprise and trivia-worthiness (One-Tailed Fisher Exact Probability Test, \(p < 10^{-50} \)).
%\dnote{What test did you run? I'm getting much much smaller p-values}
%\dmtsurel{Changed it to 0.0001. We could add more if necessary.}
%\dnote{when reporting stats, always say which test you ran}
%\ig{also report the value of $r$ for correlation, not just $p$ - should be high, i.e, close to $1$}

\begin{table}[h]
\centering
\caption{\small Trivia-worthiness and surprise}
\label{triviaSurprise}
\begin{tabular}{ccc}
 & Surprising & Not Surprising \\
Trivia-worthy & 102 & 22 \\
Not Trivia-worthy & 12 & 250
\end{tabular}
\end{table}

%\ig{perhaps worth reporting, for the MTurk study, the distribution of majority votes - in how many it was 3, in how many 4, and in how many 5} \dmtsurel{Tested. the results are in our favor, but the p-value is .052, is it worth writing about?}\dnote{what's the effect size? As in, what are the results?} \dmtsurel{Here's the agreement table \ref{Agreement} for facts that were rated as good trivia:}
%\dnote{I replaced the new table with a paragraph in the text}

\remove{
\begin{table}[h]
\centering
\caption{Agreement}
\label{Agreement}
\begin{tabular}{llll}
    & 3  & 4  & 5  \\
WTM & 21 & 16 & 5  \\
Top & 23 & 18 & 20
\end{tabular}
\end{table}
}

\subsection{Engagement Study}

In addition to the direct approach used in the first study, we conducted an additional study to indirectly measure how engaging trivia facts are.

\begin{figure}[t]
\centering
\includegraphics[width=0.7\linewidth]{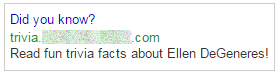}
\caption{\small An example ad used in the engagement study.\label{fig:ad}}
\end{figure}

In this study, we were targeting users who searched the entities in our dataset on the Web. We used Google AdWords~\cite{goodman2005winning} to buy ads pertaining to these entities (see Figure~\ref{fig:ad}). When users clicked an ad, they were directed to one of three variations of a landing page. The variations corresponded to the {\bf Top Trivia}, {\bf Bottom Trivia} and {\bf WTM} algorithms. Note that the ad itself was the same for all three conditions. Furthermore, we turned off Google's optimization algorithms, to ensure that the users would be uniformly distributed between the conditions.

\begin{figure}[t]
\centering
\includegraphics[width=0.78\linewidth]{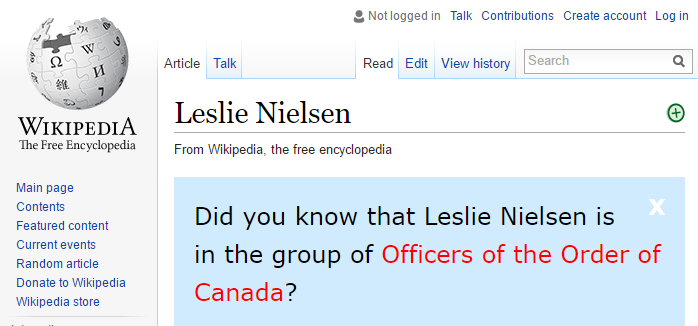} \\ 
{\LARGE $\cdot\cdot\cdot$ }
 
\includegraphics[width=0.77\linewidth]{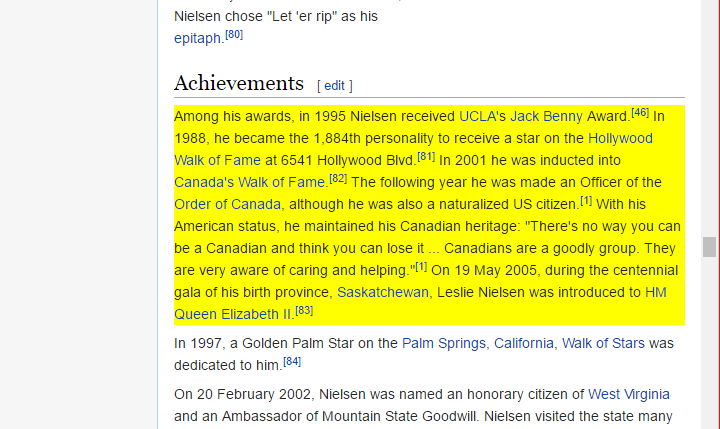} 
\caption{\small A partial screenshot of a landing page. It is a mirror of the Wikipedia page, with the trivia fact at the beginning (top) and the corresponding parts of the article highlighted (bottom). \label{fig:ln}}
\end{figure}

Each page began with the trivia fact extracted by the corresponding algorithm, and then provided a mirror of the original Wikipedia article, with the fact highlighted (Figure \ref{fig:ln}). Users were directed to paragraphs in the article text that contained broader context. For category-based facts, these paragraphs were chosen as those with the highest \artsim\ value, compared to the category title. ``Click here for another random fact'' allowed users to navigate to other trivia pages generated by the same algorithm. We conjectured that better trivia facts will engage the users more.

%To measure each algorithm separately, different sites were maintained for the different algorithms.
%\dmtsurel{Should we link to trivia.butteredcatlabs.com?}
%\dnote{no reason}

% \dnote{So basically they note CTR and bounce rate as the key measures for ad success and we can claim that since CTR was irrelevant in our case, we inspected bounce rate and then dwell time as another attempt (but maybe put more emphasis on bounce rate... do we have statistics significance for these?}

% Therefore, for each of the three variations, we measured the average dwell time of users on the site, across users who clicked the ad. 

\xhdr{Results}
We collected nearly 500 clicks throughout the experiment. Key measures for ad success
are click-through rate and bounce rate \cite{sculley2009predicting}. We also measured average dwell time of users on the site.

CTR is the percentage of users clicking on an ad. We use CTR to gauge users' level of interest in trivia in a real-life search scenario.  
In our study, CTR was inconsistent over different days, but overall averaged to $0.8\%$. Baseline CTR values in search ads is considered commercially sensitive information, so it is difficult to find non-normalized reference points. According to a recent analysis~\cite{CTR}, this value does indicate willingness of users to explore trivia facts.

%\dnote{Ido, Dan -- closing sentence. are we happy? cite doubleclick? do we have anything more scientific?} 

Next, we compared our three groups of users. 
For the {Bottom Trivia} condition, $52\%$ of users bounced immediately out of the site (under 5 seconds). WTM had $47\%$, and Top Trivia $37\%$. 
Some of the bounce rate might be explained by misguided clicks. For example, an Ellen DeGeneres ad was shown to people whose search included ``Ellen'', and was clicked on by people who searched for other Ellens.

Average time on the site for the users who did not bounce was $30.7$ seconds for Bottom Trivia, $43.1$ seconds for WTM and $48.5$ seconds for Top Trivia.
We used the one-sided Mann-Whitney U test to test whether that difference was significant.
Our hypothesis was that {Top Trivia} users stayed longer on the site.
{Top Trivia} was indeed better than {Bottom Trivia} ($p \approx 0.02$). However, the difference from {WTM} was not statistically significant ($p \approx 0.15$).

%\ig{I guess the average dwell time includes the bounce cases? So would it be fair to assume that for those who did not bounce immediately, the difference was even smaller, or even in favor of WTM. I wonder if we should be upfront about this since a reviewer may notice and complain}
%\dnote{Ido -- not including bounce, clarified}

We note that dwell time is a coarse measure for trivia quality, and there are other reasons that could explain a longer dwell time. For example, {WTM} facts were often long sentences (``Mandel has mysophobia (a pathological fear of contamination/germs) to the point that he does not shake hands with anyone, including enthusiastic contestants on Deal or No Deal, unless he is wearing latex gloves''). Other times, the {WTM} facts were somewhat cryptic (Andy Kaufman's fact was ``Keep that in mind when you call''), as can also be seen by the number of people who could not understand them in the Mechanical Turk experiment (Figure~\ref{fig:MTurk_good_trivia}). Both those reasons might prompt people to spend more time on the page -- either processing longer sentences, or scrolling down to understand the context of obscure sentences. 
%\dnote{might be worth checking out if they stayed longer on pages that actually had this -- will check}

%\dmtsurel{results: column graph of time spent. average session time (if meaningful). correlation between time and results from the MTurk study.}

%\dnote{I wrote the stats script -- waiting a few more days to get more people}

%% file: 5-discussion.tex
\section{Discussion and Future Work}

In the following section, we discuss our algorithm's limitations and potential extensions. 

\xhdr{Limitations}
We note that the proposed algorithm works well for human entities. However, there are domains where categories do not include many interesting trivia facts, such as movies or cities. (For example, consider the London categories:
London, British capitals, Capitals in Europe, Populated
places established in the 1st century, Port cities and
towns in England, Staple ports)

In these domains, our algorithm's ability to find good trivia facts is limited. Note that even in such domains, false positives can generally be avoided by a $trivia$ score threshold, as the categories in these cases are homogeneous and have $trivia$ values close to 1.
%\dnote{would we know it? wouldn't everything just have <1 score?} 
%\dmtsurel{done}

We note that our algorithm currently generates only a single-template mold (``$X$ is a member of group $Y$'') that does not appeal to users. A possible direction towards breaking the template would be to return from categories to natural language sentences: given the title of a category $C$ we attempt to find in the article text a sentence $S$ that contains similar information, using a variant of the \(\artcatsim\) function. Testing this function on the ``Grammy Award winners'' category for Barack Obama gave the following sentence as the top result:
``Obama won Best Spoken Word Album Grammy Awards for abridged audiobook versions of Dreams from My Father in February 2006 and for The Audacity of Hope in February 2008.''

A related problem is that of turning trivia facts into trivia \emph{questions}. For example, ``Barack Obama won a Grammy award'' is a good trivia fact, but turning it into a question is not straightforward. ``Who won a Grammy award?'' or ``What did Barack Obama win?'' are not good trivia questions, as they both have too many valid answers. One way to generate good questions would be to contrast a well-known category with the trivia-worthy category: ``Which US president is a Grammy award winner?''

\xhdr{Other Applications}
Our goal in this paper was to find the best trivia fact for a given article. However, our algorithm can be useful for other tasks as well. For example, the top category for Abraham Lincoln was ``American Vegetarians''. This is indeed surprising, but turns out to be historically false \cite{Hudak}. Detection of anomalous information could be useful in removing inaccurate claims from Wikipedia, thereby increasing its reliability.

In addition, the proposed \(surprise\) metric can also be used to detect the most surprising article for a given category -- or the least surprising one. For example, our algorithm detected that the least surprising article in the ``British television chefs'' category was Gordon Ramsay, who is indeed very prominent in that category.
%, as verified by the Google search entity cards when searching for that topic.

%\ig{do we have another example here that is not the entity itself? we already gave a similar example in the result analysis}. 
%\dnote{yeah -- let's try to say stuff like who's the most typical singer/Computer scientist/something}
%\dmtsurel{changed the example. what do you call the list of people that appear when you search for categories like ``british television chefs'' in google?}

We largely framed the trivia insertion problem as one that piggybacks on top of search. However, search today is just another function the smartphone performs. % to meet the user's needs. %People (and especially very young ones) are accustomed to speak their need, and have it handled by the mobile device, where t
The boundaries between entertainment and information are blurred %between local and remote content, and 
%In this kind of companion/butler setting, 
 (as evidenced by the increase of music video queries in voice search \cite{guy2016}). Combined with location data to help identify if the user is ready for listless exploration, the technology presented here could help build proactive educational agents.
%\dnote{the educational part kinda comes out of the blue here}

\xhdr{Extensions}
There are multiple dimensions we can add to our formulation, most important of which is probably \emph{personalization}. Bob Marley's Syrian-Jewish descent might be more interesting to people who are Syrian, Jewish (or both). A growing body of work looks into personalization in recommender systems. Mejova et al.~\cite{Mejova:2015:BUP:2702123.2702152} suggest personalized trivia facts as a method of breaking the ``Filter Bubble'' of social networks and increasing user interest in geographically remote countries. Young people and older people might enjoy different facts: in \cite{miliaraki2015}, there is a strong match between the age of the suggested person entity and the age of the searcher. The diversity of countries and cultures can create unique perspectives on what is obvious and what is surprising \cite{GamonMP14}. %Many other parameters can improve the personalization and localization quality of results.

In the absence of data about personal preferences, we can use \emph{popularity} as an aggregated signal. The number of page views can indicate general level of interest in a category. Temporal popularity patterns can be used to bias our algorithm (e.g., showing somebody's Irish descent just before Saint Patrick's Day).

%The number of page views an article gets is a good measure for its popularity. This feature could be used to retrieve articles that are not well known, and have a potential to be surprising. The majority of unpopular articles are unpopular for a reason, however - they are not interesting to the general public. Extracting those articles that are unpopular but also interesting would require careful work.

%\dnote{removed the part about doc2vec, seems kinda too general. perhaps get into related work}
\remove{
It would be interesting to compare the performance of different article similarity metrics. A vector representation of sentences, paragraphs and documents has been suggested as an alternative to the bag-of-words model~\cite{LeM14}. Controlling the resolution level for textual comparisons could optimize the results.
}

%Future work could use the insights and definitions presented in this paper and apply them to other domains.

%% file: 6-rw.tex
\section{Related Work}
%\dnote{those xhdrs are just for you -- remove them when you're done. should be clear from the text}

%There is a diverse body of work that is related to our problem. In
%the following, we briefly review some of the main related areas.

%\xhdr{Trivia}
There is relatively little work in Computer Science focusing on trivia.
In the work closest to ours, Prakash et al.~\cite{prakash2015did} introduced the WTM algorithm. This work used supervised learning to extract linguistic and entity-based features from a labeled dataset derived from the IMDb (Internet Movie Database) trivia section. 
Unlike our method, WTM algorithm does not utilize Wikipedia's structure. In addition, its application is limited to domains where large free labeled databases such as IMDb exist. WTM is used as a baseline in Section \ref{sec:eval}. Despite being simpler, our algorithm finds better trivia facts.
%The standard techniques of SVM learning treat Wikipedia only as a textual source of information, and do not utilize its structural properties to gain insights into relations between different parts of the data. Learning from a single specific domain can limit the algorithm's ability to generalize to other domains where large free labeled databases such as IMDb do not exist.

Merzbacher \cite{merzbacher2002automatic} tackled a related problem of mining trivia questions from a database. The questions are constructed by composing together functions (for example, the standard relational algebra operators). 
Serban et al.~\cite{DBLP:conf/acl/SerbanGGACCB16} applied a neural network architecture
on the Freebase knowledge base to transduce template-based relations into natural-language questions.
In contrast to our approach, these methods assume a relational database structure, and thus have limited applicability. 

%Structural fact databases are harder to generate, and cannot compare to the breadth of natural language information sources such as Wikipedia.
%\dnote{... and}

%\xhdr{Interestingness}
%\dnote{flow sentence introducing the area}
There is a large body of work devoted to the more general questions of surprise, interestingness and anomaly detection \cite{chandola2009anomaly}.
%\dnote{cite some stuff not on IJCAI}
For example, Byrne and Hunter \cite{Byrne2004265} develop a logic-based framework that translates structured news reports into formulas, identifying as interesting those that violate consistency or contradict
axiomatic beliefs and expectations. 
Gamon et al.~\cite{GamonMP14} consider the concept of interestingness as a user's desire to know more about a topic. By observing web-browsing logs of transitions between Wikipedia articles they construct a probabilistic model that learns latent semantic features that are interesting to users.
McGarry \cite{McGarry:2005:SIM:1103815.1103817} conducts a literature survey of interestingness measures used in knowledge discovery, divided into objective statistical measures and subjective measures based on  user beliefs or a specific domain.
Malone et al.~\cite{sunderland5282} define differential ratio rules to detect interesting patterns in spatio-temporal data. The technique uses ratios of features over time to detect change, similar to our definition of \emph{trivia-worthiness}.

%\dnote{That Hector/Aditya paper about interestingness/goodness metrics}

%\xhdr{Similarity}
%As a method of overcoming the weakness of the bag-of-words model in detecting semantic similarities, an alternative method proposed by Le and Mikolov~\cite{LeM14} extends the word vector representation of word2vec\cite{mikolov2013efficient} to represent variable-length texts such as sentences, paragraphs and documents as dense fixed-length text vectors.

%Other lines of work explore the notion of sere\paragraph{Serendipity in Recommender Systems}
In recent years, the importance of \emph{serendipity} as a measure for the success of recommender systems has grown, as one of the most prominent ``beyond-accuracy'' measures~\cite{McNee06being,Ge10beyond}. McNee et al.~\cite{McNee06being} define it as  the experience of getting an unexpected and fortuitous item. Desrosiers and Karypis~\cite{desrosiers11comprehensive} tie it with helping users find something interesting they might not have otherwise discovered. Herlocker et al.~\cite{Herlocker04evaluating} define serendipity as the extent to which the items are both attractive and surprising to users. Sun et al. ~\cite{DBLP:conf/icwsm/SunZM13} define serendipity in social networks context as messages  unexpected from the sender and relevant to the receiver.
Producing serendipitous recommendations is performed by various means, such as promoting items that have both a strong positive and a strong negative prediction scores~\cite{iaquinta08introducing} or items that are well connected, in a graph representation, both to the user's preferred items and to unrelated items~\cite{Onuma09tangent}. Evaluating serendipity is also a challenge. A recent study of social-stream item recommendation~\cite{guy15islands} directly asked participants if they found the recommended items surprising in order to asses serendipity. We ask a similar question about trivia facts in the user study conduced as part of our own evaluation. 

Serendipity was also explored in the context of \emph{search}. Recently, Miliarki et al.~\cite{miliaraki2015} demonstrated a search module which explores entities related to a search query. It was proven to be an effective vehicle for drawing searchers to an exploratory activity. 
%Moreover, it was shown to be personalized, along at least two dimensions. First, in the propensity to participate (one finding is that younger users do it more, while older users prefer the more traditional, focused search). Second, in the types of entities explored during activity (e.g., the age of celebrity viewed correlates well with the age of the viewer). 
Interestingly, the highest engagement was registered when the mentioned entity was a person (as compared to location, or a movie).
This serves as further motivation for our suggestion of augmenting entity search results with related trivia facts.
An issue left open is how to predict the response in advance -- that is, whether the user is ``focused'' or ``exploratory''.

\remove{
\xhdr{Engagement}
\dnote{opening sentence}
Heather L. O’Brien and Elaine G. Toms.
What is User Engagement? A Conceptual Framework for Defining
User. Engagement with Technology. Journal of the American
Society for Information Science and Technology, 59(6):938–955.

Potentially interesting: Simon Attfield, Gabriella Kazai, Mounia Lalmas,
and Benjamin Piwowarski. Towards a Science of User
Engagement. In WSDM Workshop on User Modelling for Web
Applications, 2011. ACM
}

%% file: 7-conclusions.tex
\section{Conclusions}

The prevailing view in the information-retrieval community sees the search engine as a passive librarian, looking up the facts. However, many users today expect the search engine to provide not just information, but also entertainment. We believe that with the advent of new search interfaces, it is time to re-examine the idea of adding serendipity to search. 

Building on the popularity of entities (and in particular person entities) in current search, we propose an algorithm to identify facts about people as \emph{trivia-worthy}. 
Specifically, we examine group membership in Wikipedia categories and rank them according to two dimensions: {\em surprise} and {\em cohesiveness}. Surprise relates 
to our prior on the person belonging to a given group, while cohesiveness ensures that said group is indeed interesting to begin with. We present a simple algorithm that is capable of discovering interesting facts, such as Hedy Lamarr's inventions.
%Our algorithm is capable of finding interesting facts, such as Obama's Grammy Award win, or Elvis' stint as a tank gunner.

We performed two kinds of user studies. First, directly and with crowd-sourced work, we show that our facts are judged as good trivia, surprising and previously unknown. Compared to prior work, our facts are judged as 27\% more surprising, and 45\% better trivia facts. Second, by buying ads on search pages, we show that our trivia facts attract page views with longer dwell times (12\%) and lower bounce rates (22\%), as compared to the baseline.

This application, while seemingly lighthearted, can lead to higher engagement of users searching for named entities. If successful, even a small impact on this type of queries can translate into a substantial improvement in user experience, and possibly transfer to other domains of human activity, like education.

%\dnote{this is from abstract and intro. rephrase}
%Trivia is about unusual bits of everyday knowledge; however, m